# Understanding the Transit Gap:

## A Comparative Study of On-Demand Bus Services and Urban Climate Resilience in South End, Charlotte, NC and Avondale, Chattanooga, TN

SANAZ SADAT HOSSEINI,[1] BABAK RAHIMI ARDABILI,[2] MONA AZARBAYJANI,[3]
SRINIVAS PULUGURTHA,[1] HAMED TABKHI[4]

[1]Department of Civil and Environmental Engineering, University of North Carolina at Charlotte, Charlotte, NC, USA
[2]Department of Public Policy, University of North Carolina at Charlotte, Charlotte, NC, USA
[3]School of Architecture, University of North Carolina at Charlotte, Charlotte, NC, USA
[4]Department of Electrical and Computer Engineering, University of North Carolina at Charlotte, Charlotte, NC, USA

ABSTRACT: *Urban design significantly impacts sustainability, particularly in the context of public transit efficiency and carbon emissions reduction. This study explores two neighborhoods with distinct urban designs: South End, Charlotte, NC, featuring a dynamic mixed-use urban design pattern, and Avondale, Chattanooga, TN, with a residential suburban grid layout. Using the TRANSIT-GYM tool, we assess the impact of increased bus utilization in these different urban settings on traffic and CO2 emissions. Our results highlight the critical role of urban design and planning in transit system efficiency. In South End, the mixed-use design led to more substantial emission reductions, indicating that urban layout can significantly influence public transit outcomes. Tailored strategies that consider the unique urban design elements are essential for climate resilience. Notably, doubling bus utilization decreased daily emissions by 10.18% in South End and 8.13% in Avondale, with a corresponding reduction in overall traffic. A target of 50% bus utilization saw emissions drop by 21.45% in South End and 14.50% in Avondale. At an idealistic goal of 70% bus utilization, South End and Avondale witnessed emission reductions of 37.22% and 27.80%, respectively. These insights are crucial for urban designers and policymakers in developing sustainable urban landscapes.*

**1. INTRODUCTION**

Urban design pattern significantly shapes transportation behaviors, influencing cities to either promote or hinder public transit. Pedestrian-friendly environments with walkable and bikeable designs lead to reduced car ownership and increased reliance on public transportation. Conversely, many post-World War II American cities, adopted car-dependent designs, leading to a decline in public transit use, notably bus transit. Main challenges plague US urban design and public bus transit, including urban sprawl, reduced bus ridership, rising costs, fleet limitations, inadequate bus stop coverage, and safety concerns, contributing to a shift toward private vehicles and consequently increased carbon emissions. By mid-2022, ridership had fallen to 62% of pre-pandemic levels, with some cities, like Charlotte, experiencing a 75% drop between 2014 and 2022, particularly in post-WWII car-dependent cities (Fig.1) [1-4].

Moreover, the broader availability of personal cars, driven by affordable ownership and low gas prices, contributes to a national decline in public transportation usage, leading to heightened traffic congestion, increased CO2 emissions, and negative environmental impacts, especially impacting underserved communities relying on older, high-emission vehicles. Traditional bus transit systems struggle to adapt to the dynamic nature of modern urban life. Our proposed solution integrates urban design principles with technology, envisioning a demand-responsive public bus system dynamically matching riders with available buses based on real-time demand [5].

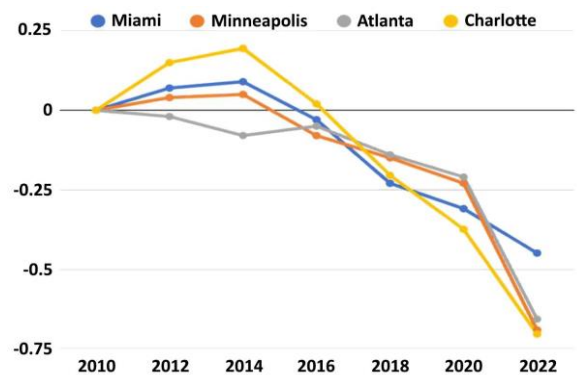

*Figure 1: Bus ridership reduction relative to 2010.*

We aim to unlock the potential of bus transit, creating an appealing ecosystem for a diverse range of commuters, including underserved communities and overlooked middle-class residents from various urban settings in cities. This study conducts a comparative analysis of two case studies, South End in Charlotte, NC and Avondale in Chattanooga, TN, utilizing the TRANSIT-GYM simulation tool to model potential impacts of this bus system with increased utilizations on areas with different urban design layouts. Preliminary results show that pedestrian-friendly areas like South End exhibit less transit inequality, with simulations indicating reduced traffic congestion and $CO_2$ emissions. Doubling bus usage reduces daily emissions in South End by 10.18% and Avondale by 8.13%, with overall traffic decreasing by 12.06% in South End and 8.90% in Avondale. At 50% bus utilization, South End sees a 21.45% drop in daily emissions, while Avondale experiences a 14.50% decrease. A 70% bus utilization target results in a 37.22% emissions reduction for South End and a 27.80% decrease for Avondale.

The outcomes of these simulations offer a roadmap for how municipalities might approach urban planning in the future, ensuring that it is both efficient and environmentally conscious. The main contributions of this paper are as follows:

1. Utilizing the TRANSIT-GYM simulation tool to quantify the benefits of increasing bus ridership through the hypothetical demand-responsive bus system and its effects on traffic reduction and $CO_2$ emissions in different urban settings.
2. Investigating how different urban design patterns affect transit behaviour, $CO_2$ emissions, and the feasibility of on-demand bus systems in South End and Avondale.
3. Highlighting the transformative potential of increasing bus ridership in enhancing climate resilience, reducing $CO_2$ emissions, and addressing transportation disparities caused by different urban design patterns.

The paper is structured as follows: Section 2 reviews research on urban design's impact on public transit and sustainability, Section 3 explains the selection of South End and Avondale for impact assessment, Section 4 presents simulation results, and Sections 5 and 6 discusses the implications and future research directions.

## 2. LITERATURE REVIEW

Recent literature emphasizes the intricate relationship between urban design and public transit efficiency to enhance urban sustainability, particularly through pedestrian-friendly designs that address the 'transit gap.' TRANSIT-GYM, developed by R. Sun et al., serves as a dynamic tool optimizing public transit operations and energy costs, facilitating efficient and equitable planning [6]. Sen et al.'s BTE-Sim offers a simulation environment for rapid modeling and optimization of public transportation systems, empowering urban planners to enhance network performance and cost-effectiveness [7].

Tao Ji et al. stress the urgency of resilient transportation infrastructure in urban areas amidst climate change challenges [8, 9]. Winkler et al.'s study critically evaluates challenges linked to reducing urban transport emissions, underscoring the necessity for comprehensive policies targeting carbon reduction in city-level emissions [10]. Jing et al.'s research delves into the intricate relationship between public transport development and carbon emissions reduction, unveiling an inverted U-shaped curve. Their findings suggest measures such as green infrastructure, government-market coordination, and energy transformation to effectively mitigate carbon emissions [11].

These studies and tools highlight the importance of simulation and AI in creating efficient, demand-responsive transit systems, improving urban resilience, lowering emissions, and informing future research on transit improvements' impacts.

Our research adds to the current body of knowledge by conducting empirical analyses and simulations in South End, Charlotte and Avondale, Chattanooga. We focus on how transitioning from private cars to public transit influences $CO_2$ emissions and transportation efficiency in these different urban settings. This study presents two key research questions that drive our investigation and findings.

1. How significantly will the scenario reduce $CO_2$ emissions and ease traffic in the case study areas?
2. How do each area's urban design and street layouts influence the simulation results?

## 3. METHODOLOGY

In this section, we explain the rationale behind selecting Charlotte and Chattanooga, discuss data sources and the simulation process, and describe the scenarios under consideration.

### 3.1 Selection of Case Study Areas

Our study, targeting South End in Charlotte, NC and Avondale in Chattanooga, TN, aims to deliver accurate and comprehensive results by examining these cities' distinct urban designs, bus ridership patterns, and their effects on transit disparities and $CO_2$ emissions. These areas were specifically chosen for their contrasting urban environments and substantial populations, providing a rich context for analysing the impact of urban design and transportation systems on $CO_2$ emissions and community dynamics.

Charlotte, NC, displays the "New South" model with its post-World War II expansion, featuring radial main streets, car-centric suburbs, and denser central areas with pedestrian-friendly grid streets.

Chattanooga, TN, shaped by its geography and history, embraces modern urban planning, like the Complete Streets Ordinance, to create a diverse environment accommodating various modes of transportation and prioritizing neighbourhood safety and interaction.

Charlotte's South End has evolved from a manufacturing district to a dynamic urban area with a blend of modern and historic elements, a comprehensive public transit system, and pedestrian-friendly spaces. Despite this, its bus usage remains low, like Avondale, and our study will analyse bus lines '5', '16', '19', '35', and '41x' [12-16].

In contrast, Avondale, a suburban Chattanooga neighborhood with mid-20$^{th}$-century roots, presents a mix of older housing in a grid street pattern, facing challenges in carpool reliance and minimal bus usage due to socioeconomic factors. Our study includes bus lines 10A, 10C, and 10G in this area [17-18].

These two neighbourhoods, each with unique characteristics, are crucial for comprehending the environmental consequences of urban planning in their respective cities.

### 3.2 Simulation and Data Sources

Utilizing the Transit-GYM tool and SUMO software [6], our study replicates traffic dynamics, transportation frameworks, and urban mobility scenarios. Multiple steps were undertaken to create SUMO scenario files for the selected areas and their bus operations, involving crafting trip definitions, establishing networks, determining vehicle configurations, and generating SUMO and GUI files. This approach aids urban transit agencies in assessing the energy implications of different transportation decisions [6]. Data for simulations, sourced from various channels [7], included:

- Map Data from Open Street Maps processed through SUMO's NETCONVERT to create SUMO-compatible road networks.
- A detailed list of public transit vehicle characteristics was compiled for SUMO's vehicle type definition file.
- The latest General Transit Feed Specification (GTFS) data (as of November 25th, 2023) provided real-time route details, bus stop locations, and schedules essential for transit simulation setup [19, 20].
- Origin-Destination (OD) data and Traffic Analysis Zones (TAZ) files were used to record trips and indicate traffic demand between zones. POLYCONVERT converted TAZ files to SUMO TAZs, outlining regions and corresponding edges, crucial for generating personal plans and vehicle trips in SUMO.

### 3.3 Bus Transit Simulation Workflow

The bus transit simulation workflow, as shown in (Fig.2), follows a structured process using SUMO tools. Initiated by collecting OD Demand matrix data, TAZ files, vehicle parameters, and GTFS, a Domain-Specific Modelling Language (DSML) program interprets these inputs, generating XML files detailing person trips, vehicle types, bus trips, and bus stop locations. Real-time interaction with the optimized network XML file and GTFS data is facilitated by the Traffic Control Interface (TraCI), which accurately positions bus stops throughout the simulation. The NETEDIT tool, a SUMO suite component, serves as an interactive graphical editor for creating and modifying road network maps for simulations. XML files representing daily non-bus traffic, integrated bus and person routes, and detailed edge-based network information are incorporated, along with average daily traffic calculations based on Annual Average Daily Traffic (AADT) and the Traffic Count Database System (TCDS) [21, 22]. These files, along with a SUMO-readable network file, amalgamate into the simulation configuration. Upon execution, the simulation assesses bus operations, providing key performance metrics such as arrival times, wait times, and load factors, offering a comprehensive view of the transport system's efficiency [6].

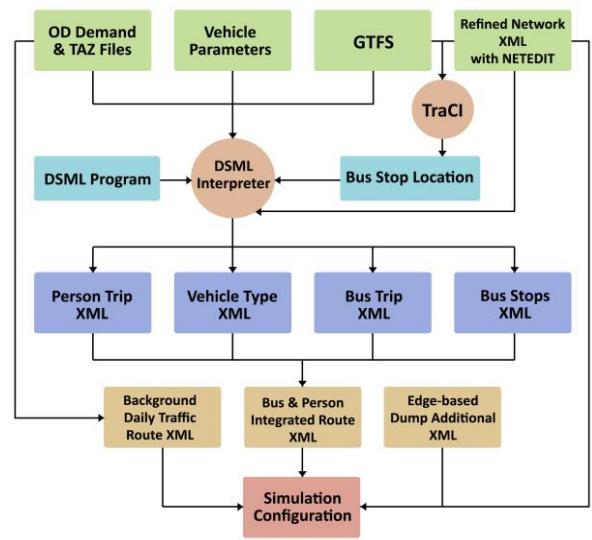

*Figure 2: Bus transit simulation workflow with TRANSIT-GYM tool [6].*

### 3.4 Scenario Selection

To assess the impact of On-Demand Bus service and bus ridership improvements on $CO_2$ emissions in two different urban areas, our study compared three scenarios against a baseline, which simulated each case study area's current street network and existing traffic patterns. The baseline reflected $CO_2$ emissions with the current bus system and ridership rate unchanged. In the first scenario, we doubled the current bus utilization within a day by transitioning private car users to a demand-responsive transit system. This resulted in a 12.06% decrease in South End and 8.90% decrease in traffic in Avondale compared to the baseline. For the second scenario, envisioning a 50% bus utilization (our proposed demand-responsive system's target), we projected a 21.13% decrease in traffic in South End and 18.40% decrease in Avondale. The third scenario assumed an idealistic 70% bus utilization, leading to approximately a 34.41% decrease in traffic in South End and 29.29% decrease in Avondale and The Results section delves into $CO_2$ emissions, examining the impact of these changes in bus trip utilization across scenarios.

## 4. RESULT

In this section, the preliminary results of the simulations under different scenarios are presented.

### 4.1 Statistical Analysis

(Table 1) compares car and bus utilization across a baseline and three increased bus utilization scenarios in South End, Charlotte. The baseline scenario captures current traffic, including the number of cars, buses, and passengers that leads to approximately 4386 reductions in car trips which increased the average bus passengers from 6.36 to 12.72. In the same scenario in Avondale, Chattanooga as shown in (Table 2) in next page, 982 unique person IDs using 173 bus line trips detected in Avondale from 5 am to 9 pm. The first scenario in Avondale, causes 660 fewer cars, which increases average bus occupancy from 5.7 to 11.40 passengers, resulting in an 8.90% decrease in traffic congestion without exceeding the current 173 bus line trips. The second scenario for both areas, leads to a significant drop in car trips: in Avondale, car trips decreased to 1363.6 trips, and in South End, from the initial number of 35335 to 28685. Also, the third scenario leads to less car trips for both areas. The necessary bus line trips for both areas remain under the existing number, indicating an enhanced bus system efficiency and no requirement for additional services.

### 4.2 Simulations Results

In this section, we represent the results of simulations in each scenario following the steps explained in the methodology section. (Fig.3) depicts the daily $CO_2$ emissions for South End, Charlotte, NC, measured from 5 AM to 9 PM across four different traffic scenarios. The base scenario peaks at 6:34 PM with total emissions of 2931.17 tonnes. When simulating Scenario 1, the peak emissions occur earlier at 5:23 PM, with a total of 2632.77 tonnes, indicating a slight decrease in emissions. Scenario 2 shifts the peak to 5:33 PM with total of 2302.41 tonnes emissions. Finally, scenario 3 leads to total of 1840.25 tonnes emissions and the peak at 7:16 PM.

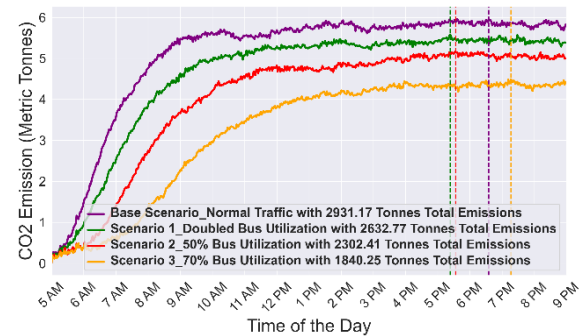

*Figure 3: Daily $CO_2$ emissions over time (hours) from 5 AM to 9 PM for different scenarios for South End, Charlotte, NC - With peak emission for base scenario at 6:34 PM, peak emission for scenario 1 at 5:23 PM, peak emission for scenario 2 at 5:33 PM, and peak emission for scenario 3 at 7:16 PM.*

(Fig.4) illustrates the daily $CO_2$ emissions in Avondale, Chattanooga, TN, from 5 AM to 9 PM across four different scenarios: the baseline, doubled bus utilization (Scenario 1), 50% bus utilization (Scenario 2), and 70% bus utilization (Scenario 3). The baseline scenario peaks at 4:55 PM with 118.72 tonnes of emissions. Scenario 1, peaks at 2:40 PM with emissions slightly lower at 109.07 tonnes. Scenario 2, peaks at 2:26 PM and the lower emissions from the above scenarios at 101.50 tonnes. Finally, scenario 3, has the lowest emission of 85.71 tonnes. The peaks show the times of day with the highest $CO_2$ emissions for each scenario. As cars are reduced and buses are used more, emissions decrease.

*Table 1: Statistical comparison of different scenarios' calculations for South End*

| Area | Scenario | Bus Utilization Data | | | Passenger Data | | Traffic Reduction Data | |
|---|---|---|---|---|---|---|---|---|
| | | Current Bus Utilization | Increase Rate of Utilization | New Bus Utilization | Current Person Trips Using Buses | Total Passengers Require Bus Services | Reduction in Car Trips | Total Average Traffic After Reduction |
| South End | Base Scenario | 18.17% | 1X | 18.17% | 6585 | 6585 | 0 | 36370 |
| | Scenario 1 | | 2X | 36.34% | | 13165.2 | 12.41% | 31983 |
| | Scenario 2 | | 3.07X | 50% | | 18112.5 | 21.75% | 28685 |
| | Scenario 3 | | 4.29X | 70% | | 25357.5 | 35.41% | 23855 |

*Table 2: Statistical comparison of different scenarios' calculations for Avondale*

| Area | Scenario | Bus Utilization Data | | | Passenger Data | | Traffic Reduction Data | |
|---|---|---|---|---|---|---|---|---|
| | | Current Bus Utilization | Increase Rate of Utilization | New Bus Utilization | Current Person Trips Using Buses | Total Passengers Require Bus Services | Reduction in Car Trips | Total Average Traffic After Reduction |
| Avondale | Base Scenario | 16.28% | 1X | 16.28% | 982 | 982 | 0 | 7412 |
| | Scenario 1 | | 2X | 32.56% | | 1972 | 9.12% | 6752 |
| | Scenario 2 | | 3.07X | 50% | | 3027.5 | 18.84% | 6048 |
| | Scenario 3 | | 4.29X | 70% | | 4238.5 | 30% | 5241 |

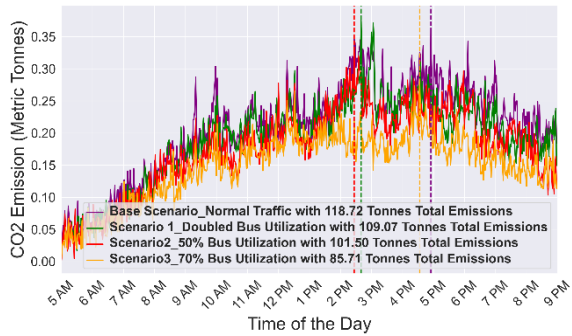

*Figure 4: Daily CO2 emissions over time (hours) from 5 AM to 9 PM for different scenarios for Avondale, Chattanooga, TN - With peak emission for base scenario at 4:55 PM, peak emission for scenario 1 at 2:40 PM, peak emission for scenario 2 at 2:26 PM, and peak emission for scenario 3 at 4:33 PM.*

The variance in emission trends between two urban areas is influenced by their distinct urban designs. In the South End, a mixed-use area, emission patterns show a steady rise from early morning, peaking around 8 AM, indicative of diverse activities starting the workday, leading to consistent emissions above 5 tonnes post 8 AM. This reflects the area's blend of residential, commercial, and leisure activities, contributing to a steady flow of traffic. In contrast, a primarily residential area shows more pronounced emission fluctuations, aligning with typical suburban patterns. These fluctuations are driven by peak commute hours, with sharp peaks during morning and evening rush hours, illustrating how urban design—whether mixed-use or residential—affects traffic patterns and, consequently, emissions.

## 5. DISCUSSION

Our study offers a comparative analysis of the Avondale and South End neighborhoods, illustrating the impact of urban design on public transport efficiency and climate resilience. South End, with its organic street network and a mix of activities, shows significant reductions in traffic and CO2 emissions, unlike Avondale's suburban grid layout. However, South End's dense pattern also leads to increased traffic congestion and emissions, as the existing street network struggles with the high volume.

In South End, the mixed-use, pedestrian-friendly environment reduces reliance on private vehicles, aiding in emission reduction. Yet, this benefit is offset by congestion issues arising from its high-density urban pattern. Avondale, with its grid pattern and residential focus, faces challenges of longer trips and higher traffic due to limited local destinations.

This research highlights the dual nature of urban design impacts on public transit and sustainability. South End's design encourages public transport use but also brings congestion challenges, indicating a need for adaptive, demand-responsive transit solutions. Avondale's experience points to the importance of incorporating local amenities in residential areas. Our findings emphasize the need for holistic urban design strategies that balance density, accessibility, and environmental impact, crucial for effective city-wide CO2 emission reduction and climate resilience.

## 6. CONCLUSION AND FUTURE WORK

The aim of this research is to illustrate how urban design affects public transit effectiveness and sustainability, in particular through the lens of demand-responsive bus services in South End, Charlotte, NC, and Avondale, Chattanooga, TN. It reveals how South End's mixed-use layout and dynamic land use enhance the benefits of increased bus utilization, resulting in greater CO2 emissions reductions than Avondale. As a result, urban design and transit solutions are intricately intertwined, advocating context-specific solutions. While the study relies on simulations and may not capture all real-world complexities, it paves the way for future empirical research in diverse urban settings to solidify these findings. From a policy standpoint, this study reinforces the need for comprehensive urban planning. In such planning, demand-responsive transit systems must be seamlessly integrated with broader urban policies, including zoning, housing, and environmental concerns. An integrated approach is needed to address broader challenges, such as social equity and environmental sustainability. The research sets the stage for multi-disciplinary exploration, emphasizing the importance of a holistic approach to urban development that aligns public transit enhancements with overarching urban development objectives, thereby fostering a sustainable, equitable, and resilient urban future.

**ACKNOWLEDGEMENTS**

We are grateful to the Charlotte Area Transit System (CATS) for providing crucial data. We thank everyone who contributed to the development of TRANSIT-GYM tool, particularly Professor Abhishek Dubey and his team at Vanderbilt University for their exceptional assistance. Our special thanks to Professor Srinivas Pulugurtha from UNC Charlotte and his laboratory for their invaluable assistance. Moreover, we appreciate the financial support provided by UNC Charlotte's School of Data Science and College of Engineering.